# Phase Behavior of Aqueous Suspension of Laponite: New Insights with Microscopic Evidence

Shweta Jatav and Yogesh M Joshi*

Department of Chemical Engineering,

Indian Institute of Technology Kanpur, INDIA

* Corresponding author, E-mail: joshi@iitk.ac.in

**Abstract**

Investigating microstructure of suspensions with particles having anisotropic shape that share complex interactions is a challenging task leading to competing claims. This work investigates phase behavior of one such system: aqueous Laponite suspension, which is highly contested in the literature, using rheological and microscopic tools. Remarkably, we observe that over a broad range of Laponite (1.4 to 4 weight %) and salt concentrations (0 to 7 mM), the system overwhelmingly demonstrates all the rheological characteristics of the sol-gel transition leading to a percolated network. Analysis of the rheological response leads to fractal dimension that primarily depends on the Laponite concentration. We also obtain the activation energy for gelation, which is observed to decrease with increase in Laponite as well as salt concentration. Significantly, the cryo-TEM images of the post-gel state clearly show presence of a percolated network formed by inter-particle bonds. The present work therefore conclusively establishes the system to be in an attractive gel state resolving a long-standing debate in the literature.

**Introduction:**

A variety of dispersions comprised of nano-particles are used in day to day life as well as in industries. Along with concentration of nano-particles, their size, shape, characteristics of their surface, and nature of solvent influence their microstructures,[1, 2] which in turn determine physical properties of the same. Particularly the shape of a nano-particle has profound effect on the resultant microstructure. Owing to high aspect ratio, suspensions of rod-like and disk/plate-like nano-particles show a very rich phase behavior at small concentrations.[3, 4] Overall greater the complexity of the nature of particles as well as the inter-particle interactions, more challenging is the characterization of a microstructure. One such system with intricate microstructures is suspensions of anisotropic nano-clays that are known to have wide ranging applications, and is a subject of investigation for over several decades.[3, 5, 6, 7, 8, 9] The primary particles of clay minerals are plate-like with the faces and the edges having different chemical composition.[7] Consequently, in an aqueous media the particles share the attractive and repulsive interactions, and self-assemble to show a very rich phase behavior. Among the various nano-clay systems, seldom any system might have attracted so much of attention and debate in the literature as the aqueous suspension of Laponite.[10, 11, 12, 13, 14] There are two contrasting proposals in the literature regarding the microstructure of this system, with one school proposing it to be an attractive gel,[13, 14] while the other claiming it to be a repulsive glass.[10, 15] In this work we address this problem by using rheological and microscopic tools. The rheological signatures supported by microscopic evidence put forth profound and unique insights into the phase behavior of Laponite suspensions, that settle a long standing debate in this field.

Laponite is 2:1 layer silicate synthetic clay. The primary particles of Laponite are fairly mono-dispersed with a disk like shape having a diameter in a range 25 – 30 nm and thickness 0.92 nm.[16] In a single disk of Laponite one octahedral magnesia layer is sandwiched by the two tetrahedral silica layers on the either side. In the octahedral layer some of the magnesium ions are isomorphically substituted by Lithium ions creating a deficiency of positive charge.[14] Consequently, the opposite faces of a particle have surplus electrons that they share with the sodium atoms. In a dry state Laponite disks are present in a form of stacks so that the sodium atoms reside in the space between the disks. Upon dispersing Laponite in aqueous media, water enters the inter disk space causing them to swell,[17] inducing dissociation of $Na^+$ ions, which diffuse out of the stacks into the bulk due to osmotic gradient. Dissociation of $Na^+$ ions renders the faces of Laponite particles a permanent negative charge. The edge of a Laponite particle is formed of broken crystals of MgO and MgOH that is known to acquire positive charge below the pH of around 11 and negative charge at higher pH.[18] Overall, below pH 11 of the suspension, Laponite particles have edge – to – face electrostatic attraction while face – to – face electrostatic repulsion. In addition, the Laponite particles also share the van der Waals attraction. The presence of salt in an aqueous media increases the ionic concentration, which shield the charges on the particles, hence influencing the electrostatic interactions among the same.[19]

Owing to the complex inter-particle interactions, aqueous suspension of Laponite has been a subject of intense investigation. Different characterization techniques such as light scattering,[10, 11, 13, 20, 21] x-ray scattering,[8, 10, 22, 23, 24] neutron scattering,[23, 24] rheology,[12, 14, 19, 22, 25, 26] microrheology,[27, 28, 29, 30, 31, 32] visual observation,[8, 10] microscopy,[22] and simulations[8, 10, 22, 32, 33, 34, 35] have been used to study a microstructure of suspension as a function of Laponite concentration ($C_L$) and salt concentration ($C_S$). Various phase diagrams have been proposed, revised and re-proposed over past two decades. It is observed that for $C_L < 1$ weight % suspension undergoes phase separation.[10, 13] Typically, above a concentration of 1 weight %, Laponite suspension shows phase transition from a free flowing liquid – immediately after preparation – to a soft solid over a period of hours to months depending upon $C_L$ and $C_S$. Higher the value of $C_L$ or $C_S$ is, smaller is the time required to form a soft solid.[10, 36] There is a consensus in the literature that for a concentration range between 1 and 2 weight %, the resultant soft solid is comprised of an attractive gel formed by the bonds between the negative faces and the positive edges.[10, 13, 15, 37] On the other hand, for $C_L$ above 2 weight %, there is no consensus regarding a microstructure responsible for a soft solid-like state. For this concentration regime, one proposal suggests that Laponite particles form a repulsive glass, wherein particles remain in a self-suspended state without touching the neighboring particles due to repulsion among the negatively charged faces.[10, 15, 33] In contrast, the other proposal indicates formation of an attractive gel with edge – to – face bonds leading to a fractal network.[13, 14, 19, 25, 26] Interestingly for a narrow Laponite concentration range of 1.8 to 2 weight %, it has been claimed that the system may evolve to either repulsive glass or attractive gel state.[20] For $C_S$ above 20 mM, flocculation is observed.[13]

The analysis of the literature suggests that the critical issue is whether microstructure of Laponite suspension for $C_L > 2$ weight % is in an attractive gel state or a repulsive glass state. Typically, distinction between a glass state and a gel state is made based on the length-scale over which the density of a material becomes homogeneous. For the glasses it is the average inter-particle distance, which is of the order of particle length-scale; while for the gels it is the average length of the strands that join the junctions in a network.[38] The scattering studies are, in principle, sufficient to distinguish between a glass and a gel state. Interestingly while the results of light scattering experiments from various groups have been interpreted to support both the types of microstructures, the characteristic lengthscale obtained from small angle X-ray scattering (SAXS) measurements have been interpreted to support the repulsive glass state for $C_L > 2$ weight %.[8, 10] The cryo-transmission electron microscopy (cryo-TEM) studies carried out on this system around two decades ago show disconnected Laponite particles in a self-suspended state for low ($C_L < 2$) as well as high ($C_L > 2$) Laponite concentration regimes.[22] The result, therefore, is still puzzling as it would then propose a repulsive glass state for the entire concentration range of Laponite suspensions.

Aqueous suspension of Laponite immediately after preparation is usually in the liquid state. The microstructure of the same evolves with time to form a soft solid thereby undergoing liquid to solid transition. Therefore, if the resultant structure is an attractive gel, suspension must undergo a sol-gel transition. On the other hand, if the resultant structure is a repulsive glass, then suspension must go through a glass transition. Interestingly rheological characterization of a material in the linear viscoelastic domain makes a clear distinction between sol-gel transition in comparison to the glass transition. The principle thesis behind the distinction is that the sol-gel transition is independent of the timescale of observation, while the glass transition is not. As a result, at the point of sol – gel transition, the damping factor given by $\tan\delta = G''/G'$, where $G'$ is the elastic modulus and $G''$ is the viscous modulus, becomes independent of frequency $\omega$, which is a measure of inverse of imposed timescale.[39, 40] At the critical gel point, $G'$ and $G''$ show an identical power law dependence on $\omega$ given by:[40]

$$G' = G'' \cot\left(n\pi/2\right) = \frac{\pi S}{2\Gamma(n)\sin\left(n\pi/2\right)}\omega^n \qquad (1)$$

where $S$ is the gel network strength, $\Gamma(n)$ is Euler gamma function defined as $\Gamma(n) = \int_0^\infty x^{n-1} \mathrm{e}^{-x} dx$, and $n$ is the relaxation exponent ($0 < n < 1$). The identical power law dependence of $G'$ and $G''$ on $\omega$ results in spectrum of relaxation times with power law distribution having a negative slope that is suggestive of a space spanning percolated fractal network.[41] Very importantly the proposed Winter-Chambon criterion[39, 40] for the critical gel state has been verified for a variety of polymeric as well as colloidal gels.[19, 42] In an important contribution, Muthukumar[43] estimated the fractal dimension of a critical gel from the relaxation exponent $n$. Very remarkably a fractal dimension obtained from Muthukumar's[43] analysis was observed to match the same obtained from the SAXS and light scattering measurements for a variety of gel forming materials.[44, 45, 46, 47]

In contrast to the critical gel transition, a point of glass transition, is characterized by a maximum in $G''$ and $\tan\delta$, and a sharp change in $G'$. A point of glass transition, however, depends on applied frequency.[48] More strikingly, the spectrum of relaxation times at the point of transition has been observed to have a positive slope, which is opposite to that observed for a gel transition.[41]

The objective of this work is to assess whether microstructure of aqueous suspension of Laponite follows characteristics of a glass transition or a sol-gel transition by studying linear viscoelastic behavior over a broad range of Laponite ($C_L$) and salt ($C_S$) concentration. In addition, we also obtain cryo-TEM images of states of some of the representative systems to directly observe the microstructure so as to complement the rheological findings.

**Materials and Method**

In this work we use Laponite XLG® (BYK Additives, Inc.; Laponite XLG® is a purer grade of Laponite RD®. Both are chemically identical in every respect) to prepare an aqueous suspension. In a typical protocol, required amount of oven dried (120°C for 4 h) Laponite XLG ($C_L$ =1.4 to 4 weight %) was mixed with ultrapure Millipore water (resistivity 18.2 MΩ-cm) having pH 10 (maintained with NaOH) and predetermined concentration of salt (NaCl: $C_S$ = 0 to 7 mM) using ULTRA-TURRAX Drive (manufactured by IKA®) for 30 min. We confirmed that suspensions remain chemical stable over the entire duration of the experiments wherein no traces of leached $Mg^{+2}$ ions have been observed. The samples thus prepared were used for rheological measurements immediately after the stirring is over. In some cases, samples were filtered using 0.45 μm (Millex®-HV, Sterile filter) filter before carrying out the experiments. The rheological experiments were performed using Dynamic Hybrid Rheometer 3 (DHR3, TA Instruments) with concentric cylinder geometry (cup diameter 30 mm and gap 1 mm). After loading a sample in the shear cell, it was subjected to cyclic frequency sweep (succession of frequency sweep experiments over the explored duration) with stress magnitude of 0.1 Pa and a frequency range of 0.5 – 25 rad/s. It takes around 2 min to complete one frequency sweep. The time $t_w$ was measured after loading a sample in the shear cell. The frequency range is chosen in such a way that the mutation numbers ( $N'_{mu} = (2\pi/\omega G')(\partial G'/\partial t)$ and $N''_{mu} = (2\pi/\omega G'')(\partial G''/\partial t)$ ) remain within limits: $N'_{mu} < 0.1$ and $N''_{mu} < N'_{mu}$, in order to have acceptable results of oscillatory experiments.[49] In this work we explore 39 systems having different $C_L$ and $C_S$. We have marked those systems by + (plus) sign on $C_L$ - $C_S$ plane in figure S1 of the supporting information. Most of the experiments were performed 2 times, which showed good reproducibility and we reported mean and error bar associated with the same. The experiments were carried out at 30°C unless otherwise mentioned. We also carry out the cyclic frequency sweep experiments at different temperatures on a system with $C_L$ =2.8 weight % and $C_S$ = 3 mM over a range: 10 – 50°C. In all the experiments we apply a thin layer of low viscosity silicon oil to the free surface to prevent evaporation.

We also report the cryo-TEM (FEI Tecnai G2 Twin TEM instrument with maximum 120 kV electron beam intensity) images for two representative samples of the Laponite suspensions. In order to take the images, filtered Laponite sample having predetermined age was mounted on a holey carbon copper grid having 400 mesh size. Subsequently the loaded grid was frozen in liquid ethane (at -188°C) using freezing reservoir of liquid nitrogen and transferred to cryoholder for imaging. We also carry out the simultaneous rheology experiments so as to link the state of a sample at the time of cryogenic freezing.

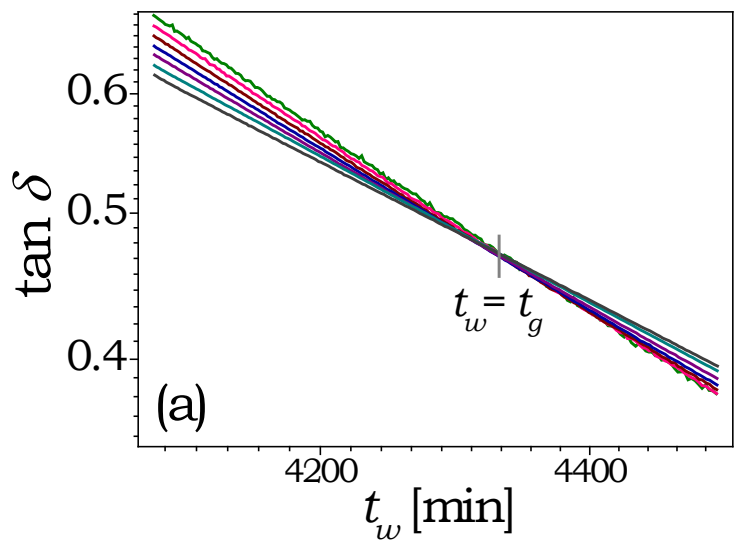

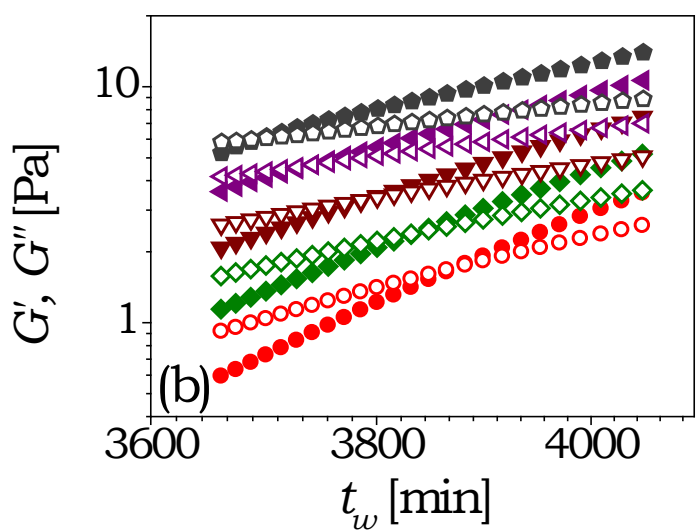

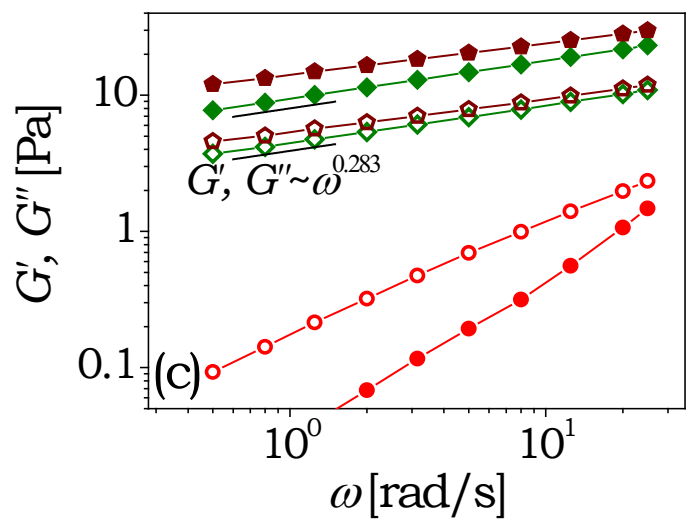

**Figure 1.** (a) Temporal evolution of $\tan\delta$ for Laponite suspension ($C_L$ =2.8 weight % and $C_S$ = 0mM) at different $\omega$ (on the left hand side from top to bottom: 2, 3.2, 5, 8, 12.5, 20, 25 rad/s). The vertical line shows the point of critical gelation. $G'$ (filled symbol) and $G''$ (open symbol) are plotted respectively, (b) as a function of $t_w$ (for different $\omega$) and (c) with $\omega$ (at different $t_w$). [For (b): $\omega$ from bottom to top 0.8, 2, 5, 12.5 and 25 rad/s and for (c): $t_w$ from bottom to top 3121, 4332 and 4492 min.] The lines in (c) serve as a guide to the eye.

Figure 1 describes the results of cyclic frequency sweep on suspension having $C_L$ =2.8 weight % and $C_S$ =0 mM. In figure 1(a) and (b) respectively, evolution of $\tan\delta$ and, $G'$ and $G''$ are plotted as a function of waiting time ($t_w$) at various frequencies. As shown, initially the suspension is in the liquid state with $G'' \gg G'$. With increase in $t_w$, both the moduli increase, however $G'$ increases at a faster rate than $G''$ resulting in $G'$ crossing over $G''$. Interestingly, $t_w$ associated with the crossover decreases with $\omega$. The figure 1(a) shows $\tan\delta$ curves at different $\omega$, which decrease with $t_w$ in such a fashion that the rate of decrease weakens with increase in $\omega$. Remarkably, all the iso-frequency $\tan\delta$ curves pass through the identical point, which is termed as the critical gel state, and the corresponding time as the gel time $t_g$. A value of $\tan\delta$ at the cross over point, given by $\tan(n\pi/2)$, leads to $n$ =0.283±0.01. In figure 1(c), we plot $G'$ and $G''$ as a function of $\omega$ for different $t_w$. At early times $G'$ and $G''$ are observed to show liquid-like behavior with $G' \sim \omega^2$ and $G'' \sim \omega$ (shown in figure S2 of the supporting information). However, with increase in time, $\omega$ dependence of both the moduli weakens, and at the point of critical gel state, both the moduli show same power law dependence on $\omega$. The corresponding critical relaxation exponent: $n$ =0.283±0.01 identically matches that obtained from the $\tan\delta$ crossover. This observation validates that the corresponding state is indeed the critical gel state with fractal network as suggested by the Winter Chambon criterion.[39, 40] At higher times, the dependence on $\omega$ continues to weaken. In figure S2 of the supporting information, we plot $\tan\delta$ as a function of $\omega$. It can be seen that at early times $\tan\delta$ decreases with increase in $\omega$, which is a characteristic feature of the liquid state. At higher times, on the other hand, $\tan\delta$ increases with increase in $\omega$, which represents a gel state. As expected from figure 1, $\tan\delta$ becomes independent of $\omega$ at the critical gel point.

We carry out the identical rheological experiments on 39 systems comprising $C_L$ in a range 1.4 to 4 weight % and $C_S$ in a range 0 to 7 mM. The corresponding dependence of $G'$, $G''$ and $\tan\delta$ on $t_w$ as well as $\omega$ for all the explored systems is plotted in the figures S3 to S8 of the supporting information. It can be seen that all the explored systems represented in figure S1 of the supporting information, show identical qualitative behavior as shown in the figure 1. This behavior clearly establishes that all the explored systems validate the rheological criterion for a critical gel state. We also carry out some experiments after filtering the sample through 0.45 $\mu$m filter. These samples also show qualitatively similar rheological behavior, which we plot in the figure S9 of the supporting information. In addition to confirm absence of wall slip, we carry out an experiment on $C_L$ =2.8 weight % with $C_S$ =3 mM suspension in a couette geometry with sand blasted surface (with identical dimensions as that of couette with smooth walls). We plot the corresponding rheological behavior in figure S10 of the supporting information. The $\tan\delta$ associated with the critical point is observed to be within experimental uncertainty with that of obtained for couette geometry with smooth walls.

In figure 2(a) we plot the time required to achieve the critical gel point ($t_g$) for all the systems as a function of $C_L$ for different $C_S$. It can be seen that gelation dynamics gets accelerated causing $t_g$ to decrease with increase in both $C_L$ as well as $C_S$. Moreover, the results suggest that $t_g$ decreases exponentially with increase $C_L$ such that on a logarithmic scale $t_g$ simply gets shifted to lower values with increase in $C_S$ without altering the nature of dependence. Consequently, shifting the experimental data vertically on the $t_g$ axis leads to superposition, which we plot in figure 2(b). The corresponding vertical shift factor $\alpha$ plotted as a function of $C_S$ in the inset of figure 2(b) shows exponential dependence. Combining dependences of $t_g$ on $C_L$ and $C_S$ yields: $t_g \sim e^{-(2.14\pm0.1)C_L-(0.84\pm0.04)C_S}$. The implications of this result are discussed below.

We plot the critical relaxation exponent $n$, which is a characteristic of the critical gel state, as a function of $C_L$ for different $C_S$ in figure 3(a). Obtaining an expression for the fractal dimension ($f_d$) for a colloidal gel of Laponite particles without knowledge of precise nature of the critical state is a challenging task. Interestingly Muthukumar[43] obtained a relationship between $n$ and $f_d$ by considering that a gel-forming system consists of a family of clusters and that the situation at the critical point is similar to arbitrarily branched polymer melt with complete screening of excluded volume effects, and is given by,

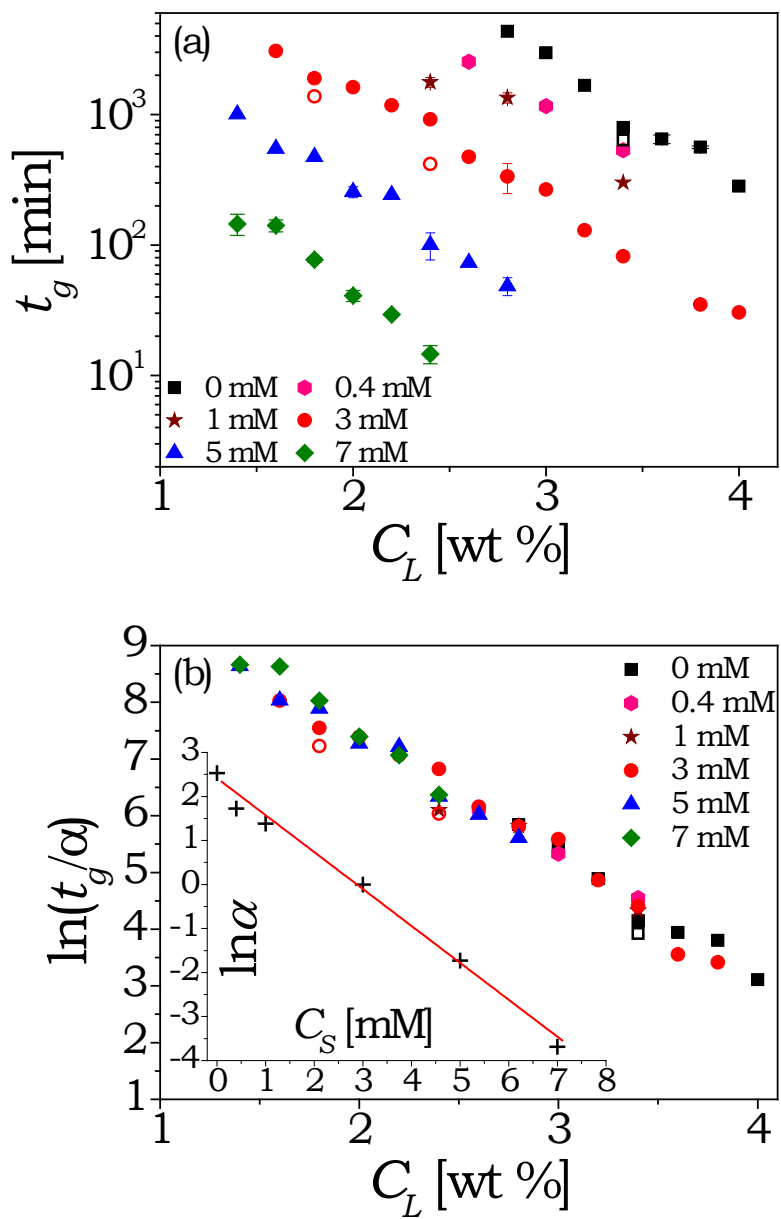

**Figure 2.** (a) The time to gel ($t_g$) is plotted as a function of Laponite concentration ($C_L$) for different salt concentrations ($C_S$). Open symbols represent filtered suspensions. (b) A superposition is obtained by shifting $t_g$ on a vertical axis using a shift factor $\alpha$. In the inset, $\ln\alpha$ is plotted as a function of $C_S$. The line in the inset represents: $\alpha \sim \exp(-0.84C_S)$.

$$f_d = \frac{(2n-d)(d+2)}{2(n-d)}, \qquad (2)$$

where $d =3$ is the dimension of space. Considering the similarity of rheological behavior of polymeric gels with that of the present system as shown in figure 1, we simply assume that the above conditions of arbitrary branched nature and complete screening of excluded volume interactions also hold for a gel formed by a Laponite suspension, and use equation (2) to obtain $f_d$. We plot $f_d$ as a function of $C_L$ for different values of $C_S$ in figure 3(b). In addition to $n$, which leads to $f_d$, we plot in figure 3(c) another important feature of the critical gel: the gel strength $(S)$ that is defined in equation (1). It can be seen that $n$ decreases while $f_d$ and $S$ increase with increase in $C_L$. Since at the critical point, $G''/G' = \tan(n\pi/2)$, depending upon whether, $G'' > G'$, $G' > G''$, or $G' = G''$, values of $n$ are respectively greater than, less than, or equal to 0.5. Accordingly, as $C_L$ increases the nature of a critical gel state changes from $G''$ dominated to $G'$ dominated. Therefore, it is not surprising that the strength of a gel $(S)$ increase with increase in $C_L$. Furthermore, since increase in $C_L$ increases the number density of Laponite particles, $f_d$ is also expected to increase as observed. Interestingly, although $n$, $f_d$ and $S$ are strong functions of $C_L$, their dependence on $C_S$ is observed to be weak. This is an important result as it suggests increase in $C_S$ increases the rate of gelation. However, $C_S$ does not affect much the nature of fractal structure as well as the gel strength.

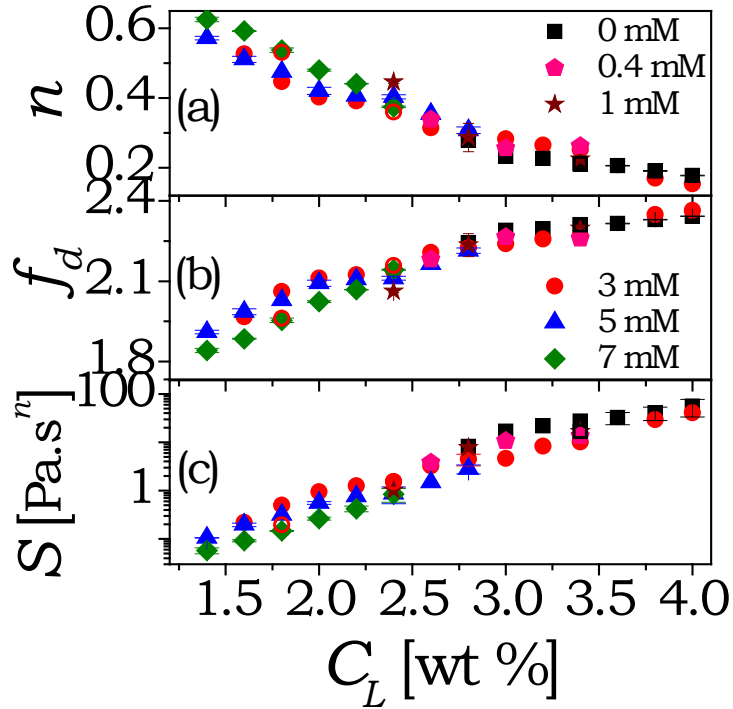

**Figure 3.** Characteristic features of a critical gel state such as: (a) the relaxation exponent $(n)$, (b) the fractal dimension $(f_d)$, and (c) the critical gel strength $(S)$ are plotted as a function of Laponite concentration $(C_L)$ for different concentrations of salt $(C_S)$. Open symbols represent the filtered suspension.

The relaxation time distribution $H(\tau)$ of a viscoelastic material can be easily obtained from $G'(\omega)$ as: $H(\tau) \approx \omega dG'/d\omega|_{1/\omega = \tau}$.[50] The representation of distribution $H(\tau)$ is such that $H(\tau)\,\mathrm{d}\ln\tau$ is the contribution to rigidity from relaxation modes whose logarithm is in between $\ln\tau$ and $\ln\tau + d\ln\tau$,[50] and for a critical gel is given by:[51]

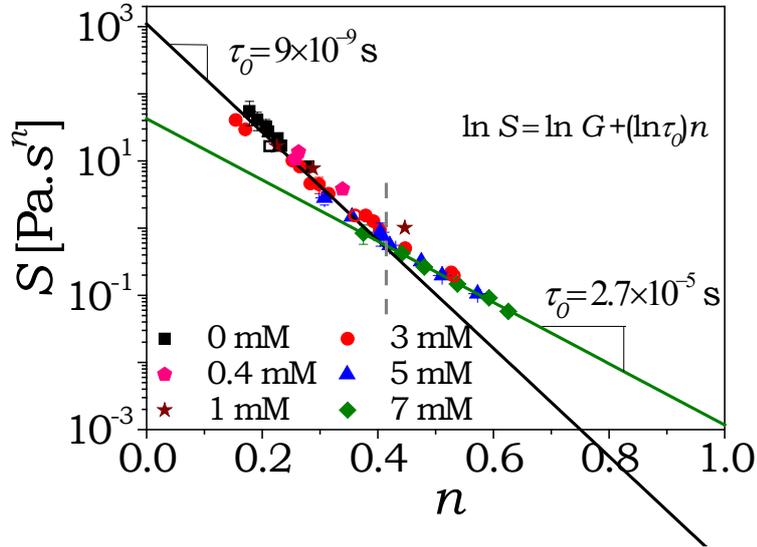

**Figure 4.** Critical gel strength $(S)$ is plotted as a function of relaxation exponent $(n)$. Open symbols represent filtered suspension.

$$H(\tau) \approx S\tau^{-n}/\Gamma(n), \qquad\qquad \tau_0 \leq \tau \leq \infty. \qquad\qquad (3)$$

Here $\tau_0$ is the lower cutoff or the shortest relaxation time associated with the power law spectrum shown by equation (3). The behavior shown in figure 3 therefore suggests that $H(\tau)$ progressively flattens as $n$ decreases for higher values of $C_L$. Furthermore, an inverse relationship between $n$ and $S$ apparent from figure 3, is routinely observed for chemical gelation.[52] The precise relation between $S$ and $n$, which renders profound insights into the nature of a network and a system, for polymeric gels is given by:[52]

$$S = G_0 \tau_0^n, \tag{4}$$

where $G_0$ is the elastic modulus of a critical gel. Unit of $S$ as manifested in equation (4) suggests that in the limits of $n = 0$ and $n = 1$, $S$ respectively represents modulus ($S = G_0$) and viscosity ($S = G_0 \tau_0$) associated with the critical gel. For polymeric gels $\tau_0$ has been observed to be the relaxation time associated with the primitive link in a network or a polymer precursor unit.[51, 52] The relaxation modes faster than this, while present in a system, do not participate in the power law spectrum.

In order to test validity of equation (4), we plot $S$ as a function of $n$ in figure 4. On a semi-logarithmic plot equation (4) describes a straight line. Interestingly, while all the data follows the same overall trend, it can be represented well by two straight lines intersecting at $n \approx 0.42$. As shown in figure 3(a), $n$ is primarily a function of $C_L$, and $n = 0.42$ represents $C_L \approx 2$ weight %. Consequently, the two straight lines - above and below this threshold - shown in figure 4 lead to $\tau_0$ associated with the Laponite suspensions in the respective domains. For $C_L < 2$, the fit of equation (4) to the experimental data leads to $\tau_0 = 2.1 \times 10^{-5}$ s, while for $C_L > 2$, the fit yields $\tau_0 = 9 \times 10^{-9}$ s. A natural characteristic timescale associated with a Laponite disk is the reciprocal of its rotational diffusivity given by $D_{r0}^{-1} = \left( \pi \eta_s d^3 \right) \big/ k_B T$, where $\eta_s$ is the solvent (water) viscosity (1 mPa.s), $k_B$ is the Boltzmann constant, $T$ is the temperature and $d$ is the diameter of a Laponite disk (30 nm). Very remarkably $\tau_0$ associated with the low concentration regime comes out to be practically identical to the $D_{r0}^{-1} = 2.03 \times 10^{-5}$ s$^{-1}$. On the other hand, $\tau_0$ associated with the high concentration regime ($C_L > 2$) is around 3 orders of magnitude smaller. It should be noted that physically $\tau_0$ suggests the lower threshold for the power law relaxation spectrum. The present experimental data therefore suggests that the lower bound of the power law spectrum continues up to inverse of rotational diffusivity of a Laponite particle for $C_L < 2$ regime. There is a possibility that that below 2 weight %, at the critical point, the smallest link in the network is formed by individual Laponite particles leading to an inverse of rotation diffusivity ($D_{r0}^{-1}$) of a Laponite particle. On the other hand, for $C_L > 2$ regime, the power law spectrum continues up to significantly lower values of the relaxation modes, wherein the smallest link could be formed by the un-exfoliated tactoids of the Laponite particles.

In a gel forming system, at the critical point, not just the viscoelastic properties show a definite behavior but also how these properties evolve as a function of time.

Particularly, for polymeric gels the rate of change of $G'$ and $G''$ is observed to follow power law dependence on $\omega$ at a critical gel point $\left( t_w = t_g \right)$ given by:[51]

$$\left( \frac{\partial \ln G'}{\partial t_w} \right)_{t_w = t_g} \approx C \left( \frac{\partial \ln G''}{\partial t_w} \right)_{t_w = t_g} \sim \omega^{-\kappa}. \qquad (5)$$

The two parameters of this dependence $\kappa$ (dynamic critical exponent) and $C$ (proportionality constant) have been observed to have universal values of $\kappa = 0.22 \pm 0.02$ and $C = 2$ for a very broad class polymeric gels irrespective of the stoichiometry, the chain length, and the concentration.[51, 53, 54, 55] The same values have also been observed for physical gels.[53] In figure S11 of the supporting information, we plot $\left( \partial \ln G'/\partial t_w \right)_{t_w = t_g}$ as well as $\left( \partial \ln G''/\partial t_w \right)_{t_w = t_g}$ as a function of $\omega$ for few systems. It can be seen that both the derivatives show similar power law dependence on $\omega$ as suggested by equation (5) that leads to the constants $\kappa$ and $C$, which we plot in figure 5. Very remarkably, for almost all the studied concentrations of Laponite as well as salt, $\kappa$ is observed to be in between 0.1 and 0.3, while $C$ is observed to be in between 2 and 5. This shows that both the critical constants are very close to what have been reported for the chemical as well as other physical gels.

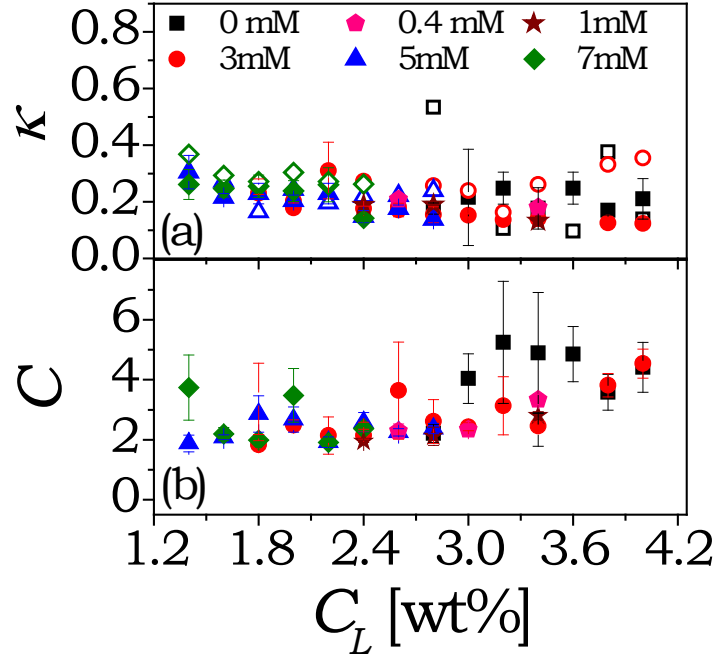

**Figure 5.** Dynamic critical exponent ($\kappa$) (a) and constant of proportionality ($C$) (b) described in equation (5) obtained from evolution of $G'$ (filled symbol) and $G''$ (open

symbol) at a critical point are plotted as a function of Laponite concentration ($C_L$) for different concentrations of salt ($C_S$).

We also perform the rheological gelation experiments at different temperatures in the range $10 - 50°$C. The corresponding evolution of $G'$, $G''$, and $\tan\delta$ are plotted in the figure S12 of the supporting information, while $t_g$, $n$, $f_d$, and $S$ are plotted as a function of $10^3/T$ in figure 6. It can be seen that $t_g$ decreases by a factor of 5 with increase in $T$ over the explored range. Furthermore, $S$ and $f_d$ can be seen to be increasing while $n$ can be seen to be decreasing with increase in $T$. Interestingly, the trend of decrease in $n$ while $f_d$ and $S$ increase with accelerated gelation kinetics (due to increase in $T$) shown in figure 6 is similar to that reported in figures 2 and 3 but with increase in $C_L$. However, for a corresponding change in $t_g$ (by a factor of 5) changes in $n$, $f_d$ and $S$ are very modest for variation in $T$ compared to that of with respect to $C_L$.

In figure 6(a), $t_g$ can be seen to be showing exponential dependence on $10^3/T$, suggesting gelation in Laponite suspension to be an activation process similar to that observed for polymer gels,[56, 57, 58] leading to the Arrhenius relationship given by:

$$t_g = t_{g0} \exp\left(E/k_B T\right). \tag{6}$$

Here $E$ is the activation energy associated with the gelation process and $t_{g0}$ is the prefactor representing gel time in the limit of $E/k_B T \to 0$. Fit of equation (6) to the experimental data as shown in Figure 6(a) leads to estimation of activation energy to be $E = 245.3$ kJ/mol (or around $97.3\,k_B T$ at $30°$C). For a gelation process, $E$ for variety of chemically gel forming polymeric systems have been reported to be in between 40 to 100 kJ/mol.[58, 59, 60] Figure 2(b) suggests that, over an explored range, dependences of $t_g$ on $C_L$ and $C_S$ is given by: $t_g \sim \exp\left[-(2.14 \pm 0.1)C_L - (0.84 \pm 0.04)C_S\right]$. Therefore, according to equation (6), $E$ decreases linearly with $C_L$ and $C_S$ given by: $E \sim -k_B T\left[(2.14 \pm 0.1)C_L + (0.84 \pm 0.04)C_S\right]$. Combining results of figures 2 and 6, and using equation (6), the specific form of activation energy in kJ/mol over the explored system variables can be obtained and is given by: $E = 266.7 \text{-} 5.4\,C_L \text{-} 2.1\,C_S$, where $C_L$ is in weight % and $C_S$ is in mM. To best of our knowledge the present work is the first to report validation of equation (6) leading to activation energy for a physical gelation process.

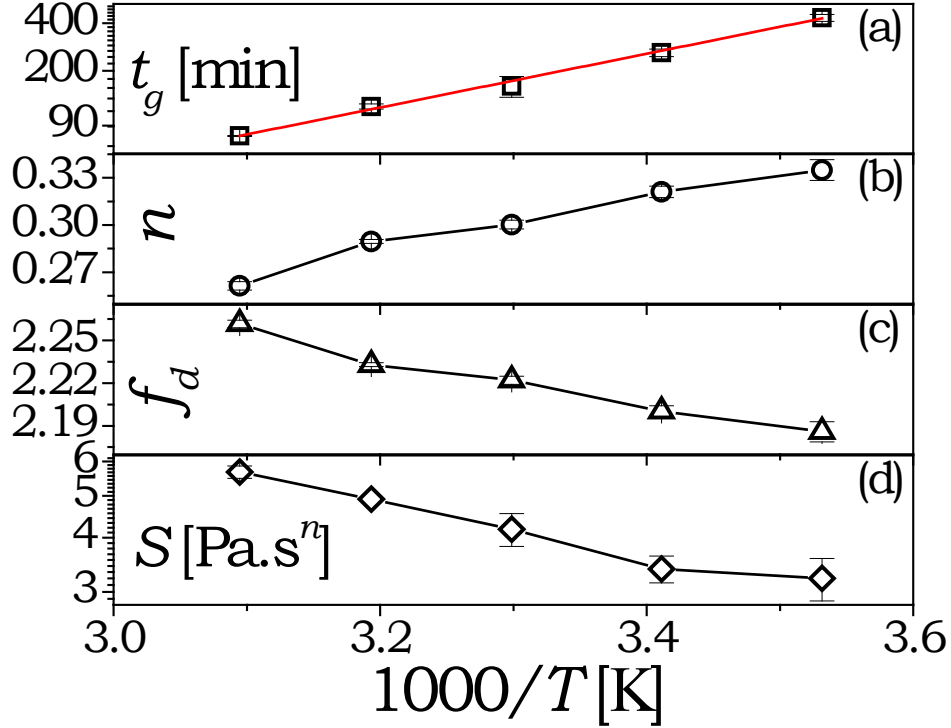

**Figure 6.** Effect of temperature on various characteristic features of critical gel state such as: (a) the time to gel ($t_g$), (b) the relaxation exponent ($n$), (c) the fractal dimension ($f_d$), and (d) the critical gel strength ($S$) of $C_L$ =2.8 weight % and $C_S$ =3 mM suspension. The lines in (a) is a fit of equation (6) to the experimental data, while lines in (b), (c) and (d) serve as a guide to the eye.

The discussion to this point very clearly indicates that the samples studied in this work that span broad range of $C_L$ and $C_S$, overwhelmingly demonstrate all the characteristic rheological signatures of critical gelation shown by the chemical as well as other physical gels. However, the systems with $C_L$ >2 weight % have been identified as repulsive glasses in the literature,[10] wherein the Laponite particles are proposed to remain self-suspended in the repulsive environment without touching each other. Therefore, in order to associate structure to the observed rheological behavior, we take the cryo-TEM images of two filtered systems: (a) $C_L$ =1.8 weight % with $C_S$ =3 mM and (b) $C_L$ =2.8 weight % with $C_S$ =0 mM. In rheological experiments systems (a) and (b) have been observed to show the critical gel states at around 32 h and 72 h respectively. For both the systems, we obtain the cryo-TEM images of the states frozen at 60 h (for (a)) and at 96 h (for (b)) of suspension's age. The corresponding images are shown in figure 7. The concentration difference between both the systems is clearly apparent from the figures 7(a) and (b). It can be clearly seen that the particles in both the figures show edge to face bonds at various angles including that of overlap coin configuration as proposed in the simulations.[35] In figure 7(a), which shows the image of system (a), interparticle bonds are clearly visible, though an explicit network is not visible. However, there is no ambiguity in the literature regarding the attractive gel microstructure of the system (a) [$C_L$ =1.8 weight % and $C_s$ =3 mM]. Therefore, it could be possible that owing to lower Laponite concentration, the 3 – dimensional percolated network structure is below the observation plane and therefore is not apparent in the image. Figure 7(b) represents a cryo-TEM image of the system (b) [$C_L$ =2.8 weight % and $C_s$ =3 mM]. It can be seen that clay particles form a very clear percolated structure in the plane of view. We believe that this figure unequivocally establishes system (b) to be in an attractive gel state.

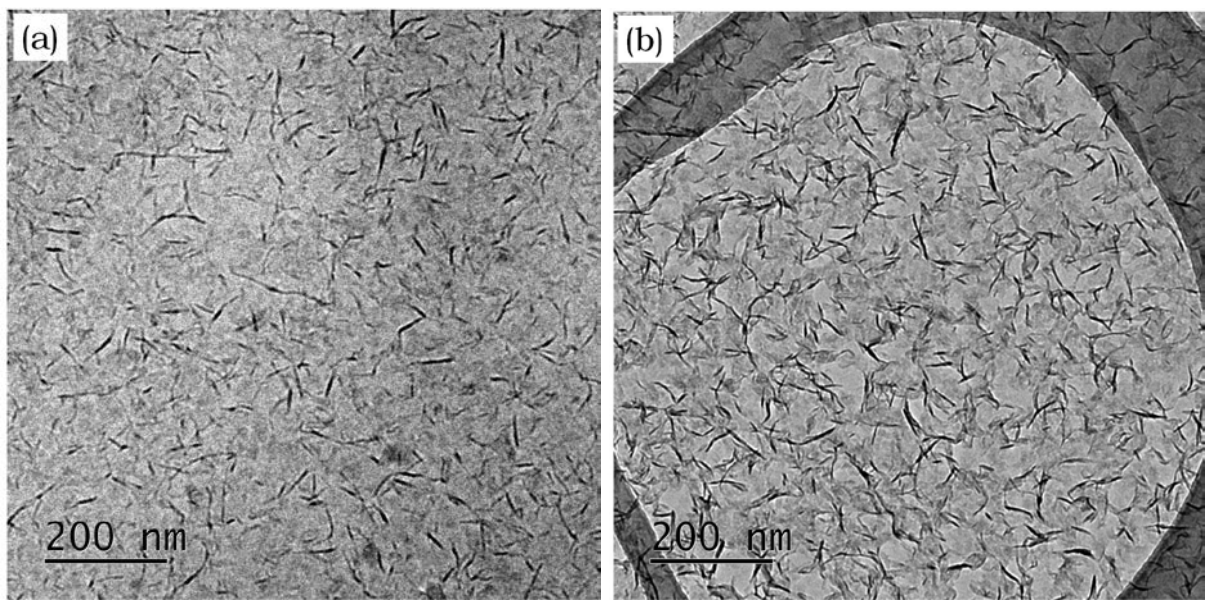

**Figure 7.** Cryo-TEM images of a post-gel state of Laponite suspensions (a) $C_L$ =1.8 weight % with $C_s$ =3 mM (age 60 h) and (b) $C_L$ =2.8 weight % with $C_s$ =0 mM (age 96 h).

The critical gel state of Laponite gel apparent in figure 1 as well as in the figures S2 to S9 of the supporting information, is associated with the power law spectrum of relaxation times given by equation (3). The power law relaxation spectrum usually represents a hierarchical fractal structure devoid of any characteristic length-scale or timescale.[61] The plot of $S$ versus $n$ shown in figure 4, however leads to $\tau_0$, which is the lowest time-scale associated with the hierarchical structure. Figures 4 and 7 suggest that rheological response and the nature of the gel state apparent from the images below and above 2 weight % concentration of Laponite suspension is different. For $C_L$ <2 weight % the cryo-TEM image shows a lean gel wherein segments in a gel appear to be formed by single Laponite particles/layers. It is therefore interesting to see from figure 4 that the smallest timescale in the spectrum ($\tau_0$) is the inverse of rotational diffusivity of a single Laponite particle. For $C_L$ >2 weight %, on the other hand, the particles appear to be present in the form of tactoids (multiple unexfoliated particle layers). For this regime of concentrations $\tau_0$ associated with the hierarchical structure extends to far lower timescales as shown in figure 4. This suggests that timescales faster than inverse of rotational diffusivity of a single particle, belonging to the unexfoliated tactoidal structure participate in the power law relaxation spectrum related to the critical gel state. Interestingly, in the literature $C_L \approx 2$ weight % has been termed as a boundary that separates an attractive gel from a repulsive glass, as physical behavior of aqueous Laponite suspension belonging to these two regimes show differences.[10, 15] The present work however, very clearly shows that, despite the differences, material in both the regimes is unequivocally in an attractive gel state.

Another important aspect associated with the Laponite suspension is time dependency associated with its structural evolution, wherein structure of the same keeps on evolving even after the sol-gel transition. It should be noted that various characteristic features including the fractal dimension discussed in this work strictly belong to the critical gel state. At higher times, the structure of the suspension that evolves to explore lower free energy states may be different from the critical gel state, however the building block of the same still remains the attractive particle – particle bonds. Jatav and Joshi[26] reported dependence of $G'$ and $G''$ on $\omega$ and corresponding relaxation time spectra beyond the critical gel state for 2.8 weight % Laponite suspension with 3 mM NaCl. The results showed that the dominance of slow modes gradually increases with increase in time. We propose that this behavior originates due to the consolidation of gel wherein on one hand density of network increases, on the other hand strengthening of the interparticle bonds take place.

Interestingly, Jabbari-Farouji *et al.*[28] report bulk as well as microrheology results for low (0.8 weight % with 6 mM NaCl) and high (3.2 weight %) Laponite concentration suspensions that very clearly demonstrate the critical gel state wherein both $G'$ and $G''$ have been observed to show the identical power law dependence on $\omega$. This confirms both the systems to be in the fractal gel state as per the Winter-Chambon criterion.[39] At higher waiting times two power law regimes of $G'$ dependence on $\omega$ have been reported. However, both the regimes constitute positive power law exponents suggesting overall dominance of the fast timescales in comparison with the slow timescales indicating a gel state in accordance with the picture proposed by Winter.[41] Very interestingly, for 3.2 weight % suspension, the displacement power spectral densities of 0.5 μm diameter tracer particle is observed to be invariant of a bead position. However, our cryo-TEM image clearly suggests that over 0.5 μm length-scale the microstructure of colloidal gel is homogeneous for even smaller concentration suspension (2.8 weight %) explaining the observed behavior very well. Finally, it is important to note that the rheological analysis presented in this work very rightly points towards the percolated network structure, which is validated by the cryo-TEM images. Ability of the rheological tools to distinguish gel structures have already been vindicated by the scattering studies for chemical gels.[44, 45, 46, 47] This work, therefore, shows rheology to be a consistent tool to characterize the physical gels as well.

**Conclusion**

This work revisits the phase behavior of aqueous suspension of Laponite, a system extensively studied and debated in the literature. We observe that temporal evolution of viscoelastic properties of aqueous Laponite suspensions, over a broad range of Laponite and salt concentrations, overwhelmingly show all the characteristic rheological features that are well established for chemically crosslinked polymeric fractal gels. Interestingly, for Laponite concentration above 2 weight %, this system has been reported to form a repulsive glass in the recent literature.[10] Our results, which indicate presence of percolated attractive gel, show that fractal dimension of a Laponite suspension gel increases with increase in Laponite concentration. The activation energy associated with the gelation process, which is observed to be higher than that observed for chemical gelation, decreases with increase in Laponite as well as salt concentration. We also obtain cryo-TEM images of Laponite suspensions in a post-gel state, which unambiguously show a network structure formed by inter-particle attractive bonds between the negatively charged faces and the positively charged edges, thereby supporting our rheological findings. This work, therefore, on one hand endorses rheology as a reliable tool to decide gel structure not just for the chemical but also for the physically crosslinked gels. On the other hand, very importantly, we believe that the present work resolves the mystery of phase behavior of aqueous suspension of Laponite by concluding it to be in the attractive gel state over the explored regime.

**Acknowledgment**

We acknowledge financial support from the department of atomic energy – science research council (DAE-SRC), Government of India. We also acknowledge advanced imaging center at IIT Kanpur for cryo-TEM facility.

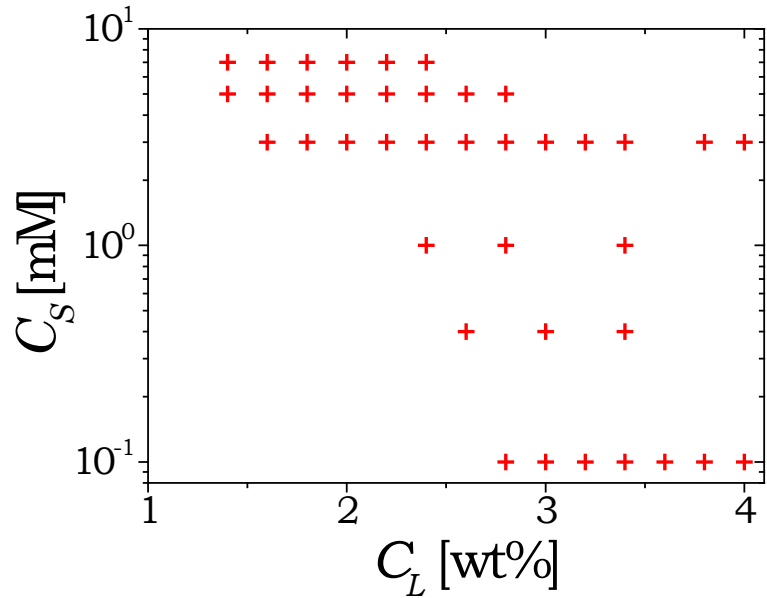

**Figure S1.** The Laponite suspension samples studied in the present work are described by + (plus) symbols on $C_L$ - $C_S$ plane. Interestingly all the samples studied in this work demonstrate various rheological characteristic features of an attractive gel with a fractal network structure.

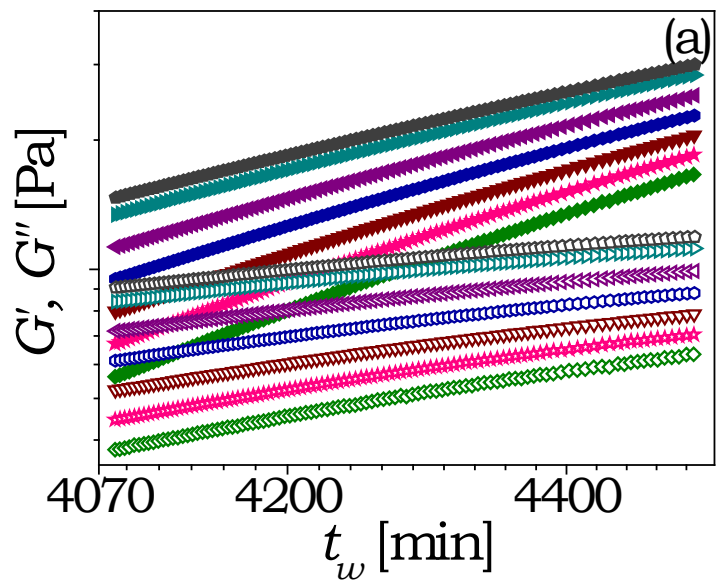

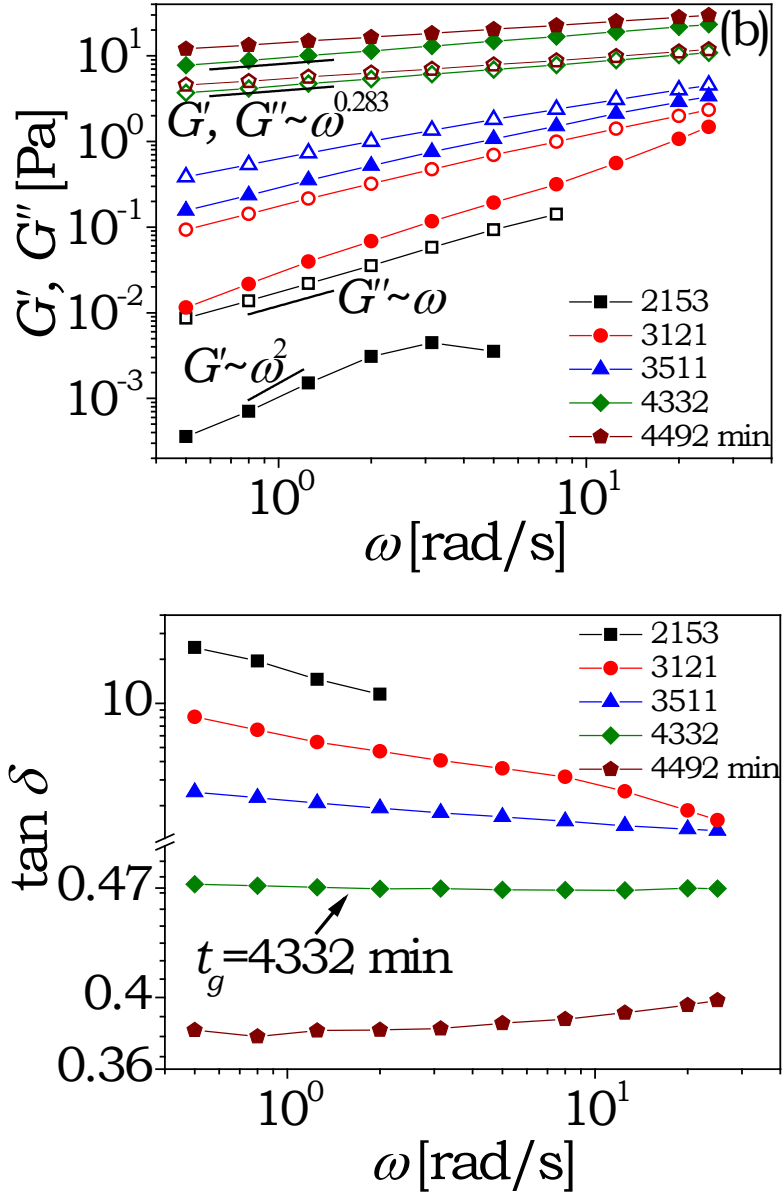

**Figure S2.** (a) Viscoelastic moduli $G'$ (filled symbols), $G''$ (open symbols) of Laponite suspension ($C_L$ =2.8 wt % and $C_S$ = 0mM) are plotted as a function of waiting time ($t_w$) for different frequencies (from bottom to top 2, 3.2, 5, 8, 12.5, 20 and 25 rad/s. Due to fluctuations the low frequency data is not shown. (b) $G'$ (filled symbols) and $G''$ (open symbols) and (c) $\tan\delta$ are plotted as a function of frequency at different $t_w$. The lines connecting the data are guide to the eye.

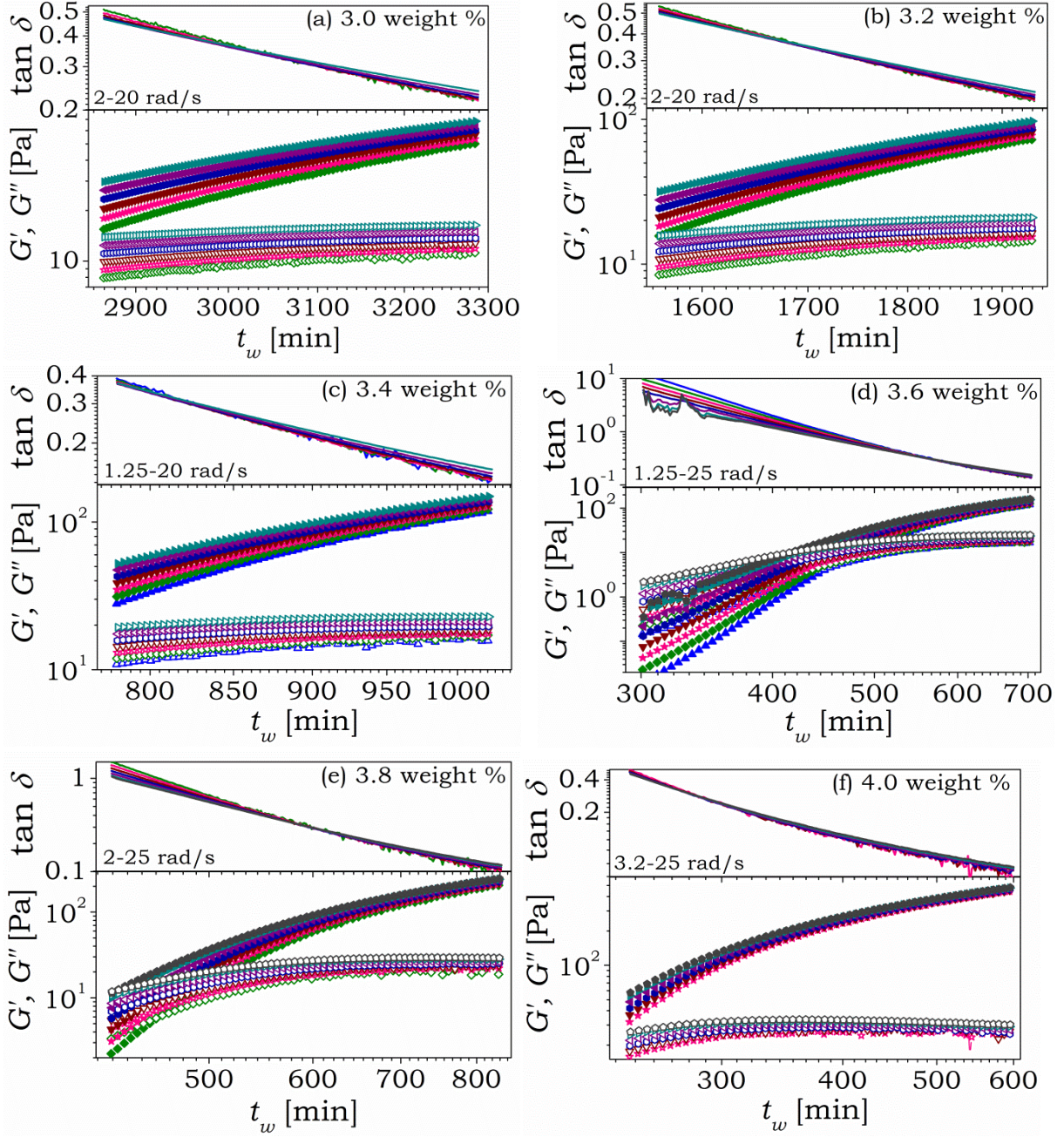

**Figure S3.** Viscoelastic properties $G'$ (filled symbols), $G''$ (open symbols) and $\tan\delta$ for suspensions of $C_S = 0$ mM and $C_L = 3.0$ (a), 3.2 (b), 3.4 (c), 3.6 (d), 3.8 (e), and 3.8 wt % (f) are plotted as a function of $t_w$ at different frequencies 1.25, 2, 3.2, 5, 8, 12.5, 20 and 25 rad/s (for $G'$ and $G''$ from bottom to top and for $\tan\delta$ on the left hand side from top to bottom). Due to fluctuations, few frequencies data are not shown for better clarity.

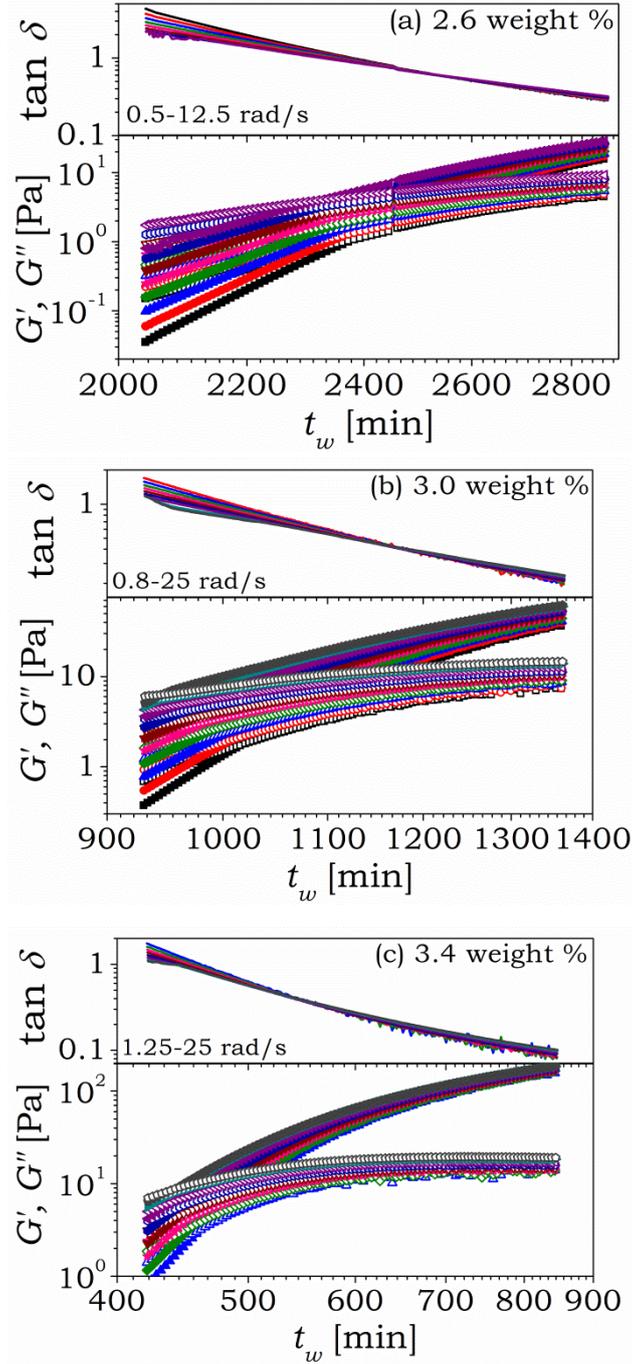

**Figure S4.** Viscoelastic properties $G'$ (filled symbols), $G''$ (open symbols) and $\tan\delta$ for suspensions of $C_s = 0.4$ mM and $C_L = 2.6$ (a), 3.0 (b), and 3.4 wt % (c) are plotted as a function of $t_w$ at different frequencies 0.5, 0.8, 1.25, 2, 3.2, 5, 8, 12.5, 20 and 25 rad/s (for $G'$ and $G''$ from bottom to top and for $\tan\delta$ on the left hand side from top to bottom). Due to fluctuations, few frequencies data are not shown for better clarity.

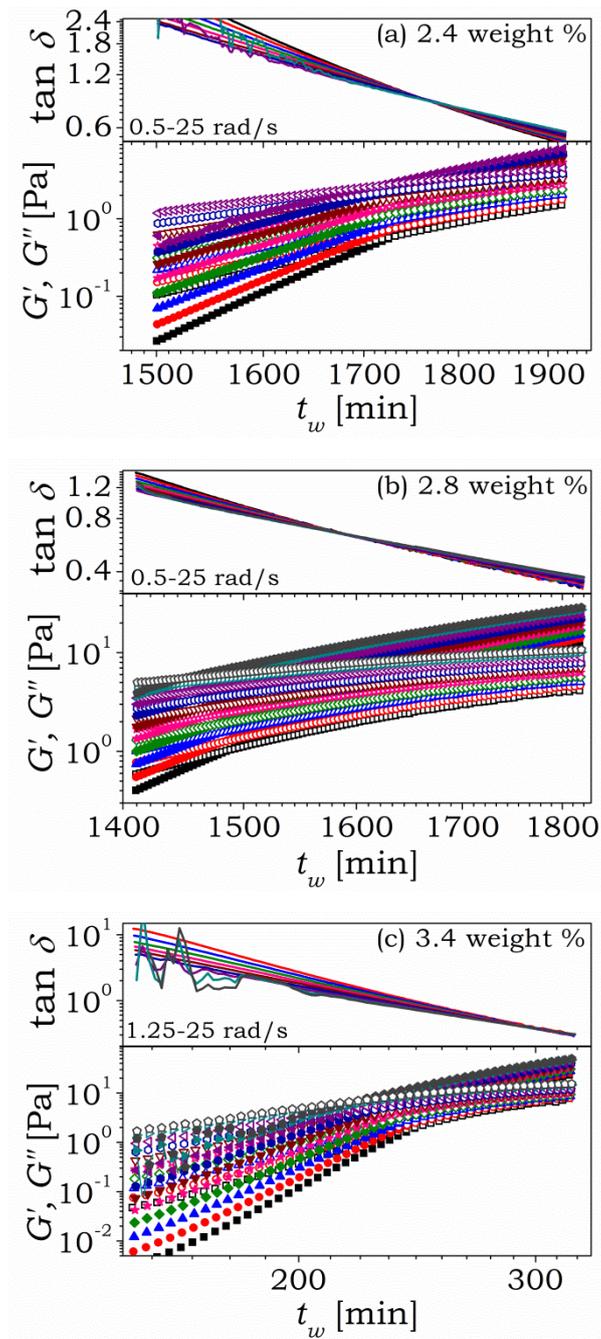

**Figure S5.** Viscoelastic properties $G'$ (filled symbols), $G''$ (open symbols) and $\tan\delta$ for suspensions of $C_S$ =1 mM and $C_L$ = 2.4 (a), 2.8 (b), and 3.4 wt % (c) are plotted as a function of $t_w$ at different frequencies 0.5, 0.8, 1.25, 2, 3.2, 5, 8, 12.5, 20 and 25 rad/s (for $G'$ and $G''$ from bottom to top and for $\tan\delta$ on the left hand side from top to bottom). Due to fluctuations, low frequency data is not shown in the last part for better clarity.

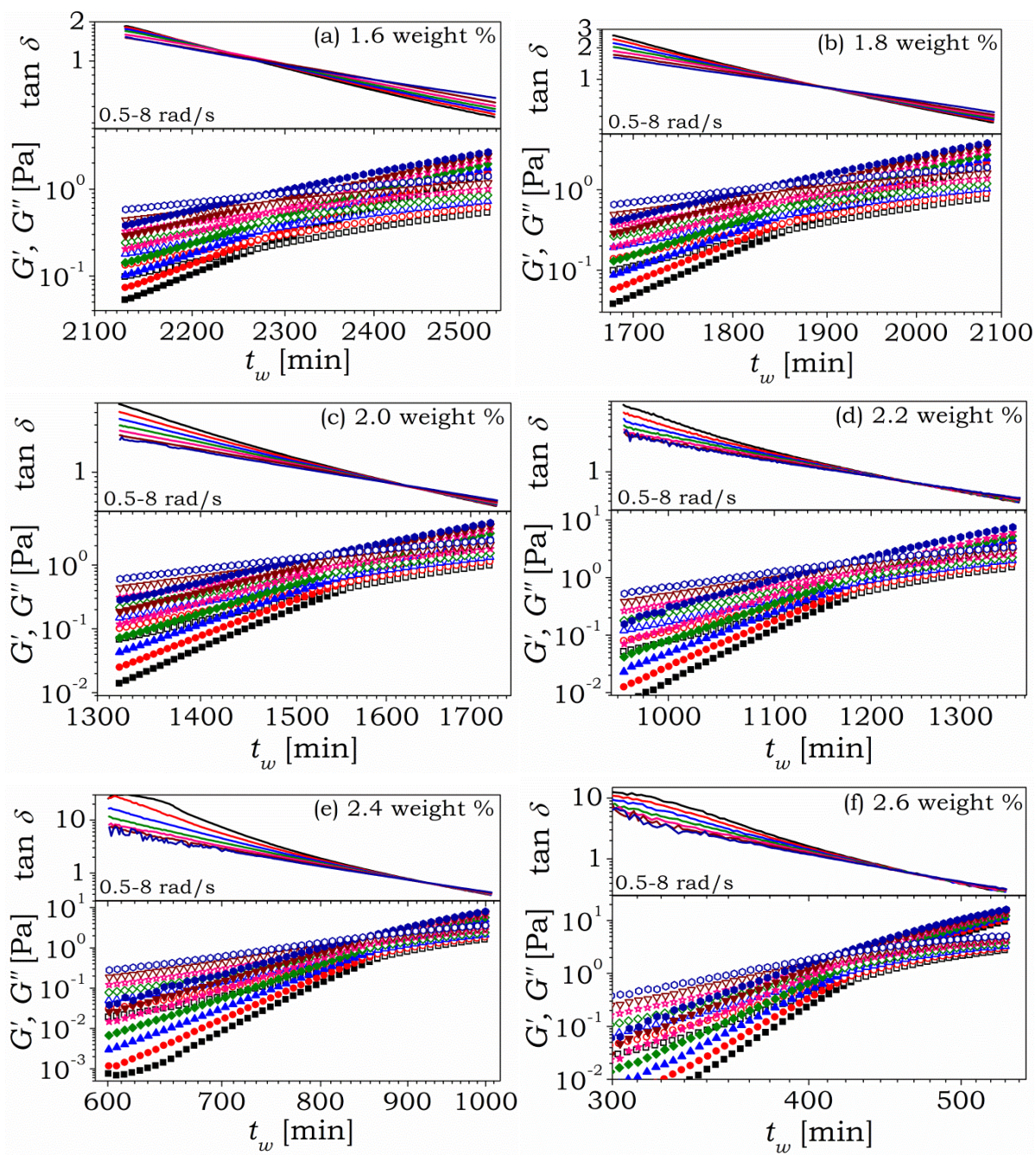

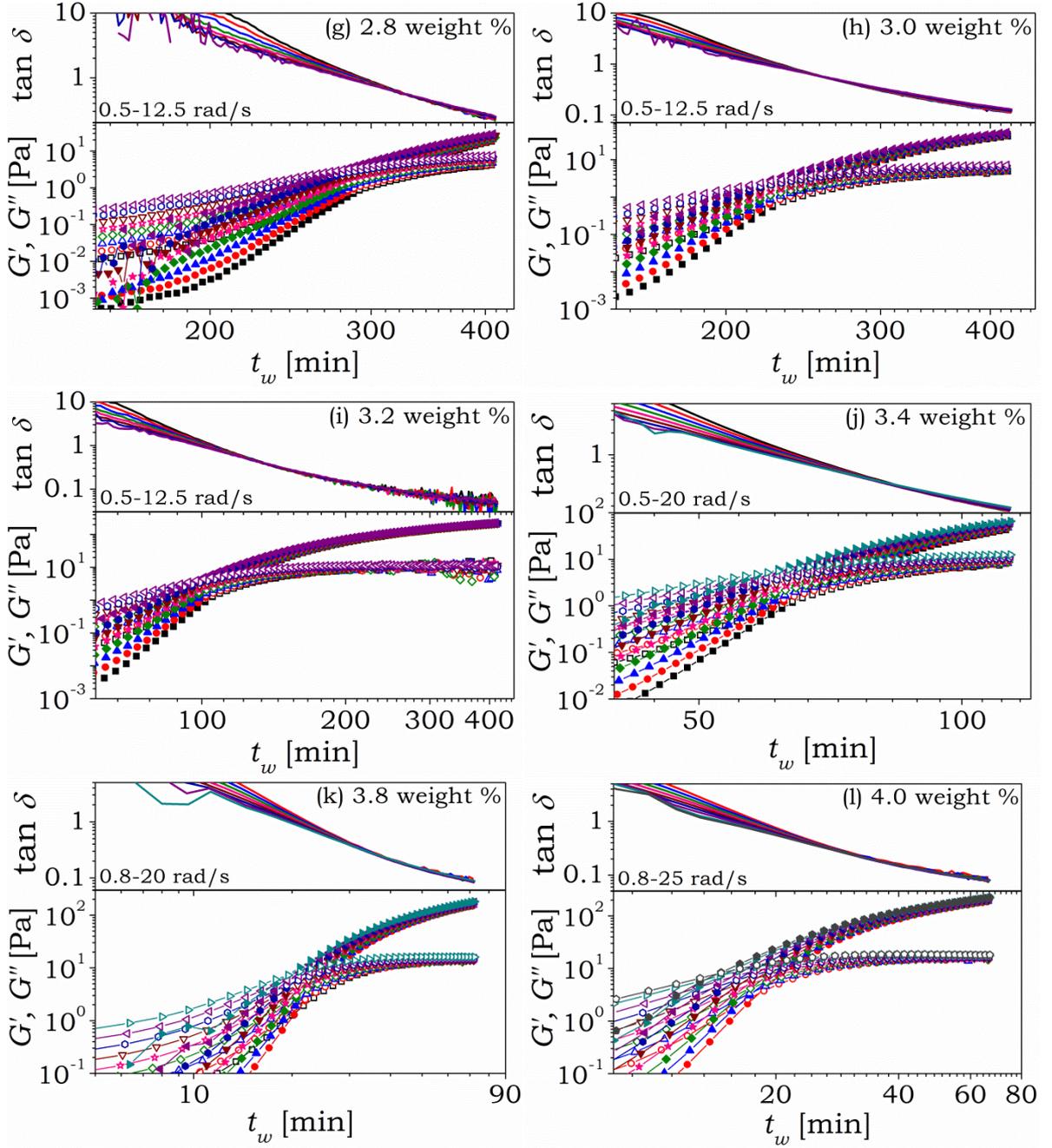

**Figure S6.** Viscoelastic properties $G'$ (filled symbols), $G''$ (open symbols) and $\tan\delta$ for suspensions of $C_s$ =3 mM and $C_L$ = 1.6 (a), 1.8 (b), 2.0 (c), 2.2 (d), 2.4 (e), 2.6 (f), 2.8 (g), 3.0 (h), 3.2 (i), 3.4 (j), 3.8 (k), and 4.0 wt % (l) are plotted as a function of $t_w$ at different frequencies 0.5, 0.8, 1.25, 2, 3.2, 5, 8, 12.5, 20 and 25 rad/s (for $G'$ and $G''$ from bottom to top and for $\tan\delta$ on the left hand side from top to bottom). Due to fluctuations, few frequencies data are not shown for better clarity.

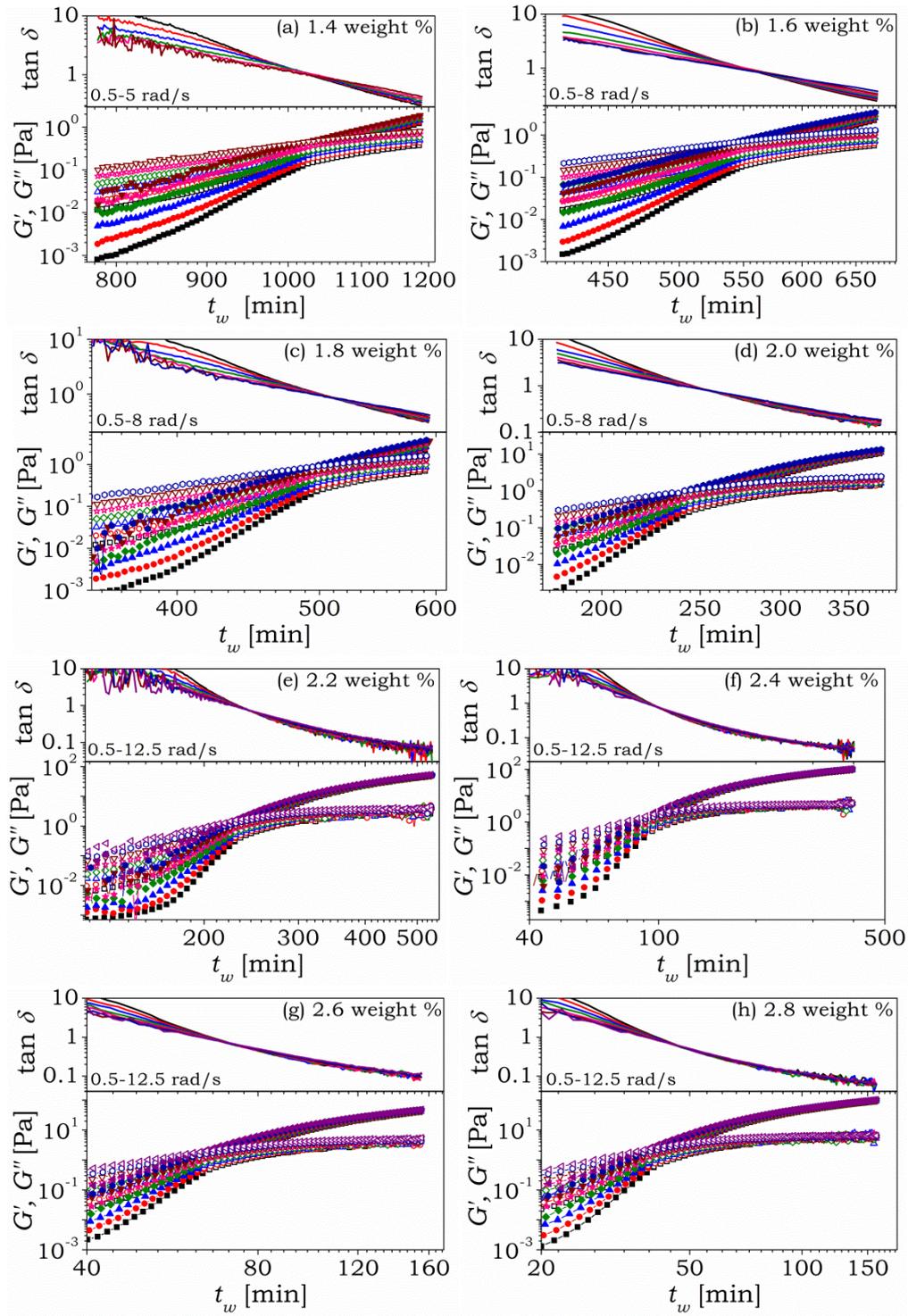

**Figure S7.** Viscoelastic properties $G'$ (filled symbols), $G''$ (open symbols) and $\tan\delta$ for suspensions of $C_s$ =5 mM and $C_L$ = 1.4 (a), 1.6 (b), 1.8 (c), 2.0 (d), 2.2 (e), 2.4 (f), 2.6 (g), and 2.8 wt % (h) are plotted as a function of $t_w$ at different frequencies 0.5, 0.8, 1.25, 2, 3.2, 5, 8, and 12.5 rad/s (for $G'$ and $G''$ from bottom to top and for $\tan\delta$ on the left hand side from top to bottom). Due to fluctuations, few frequencies data are not shown for better clarity.

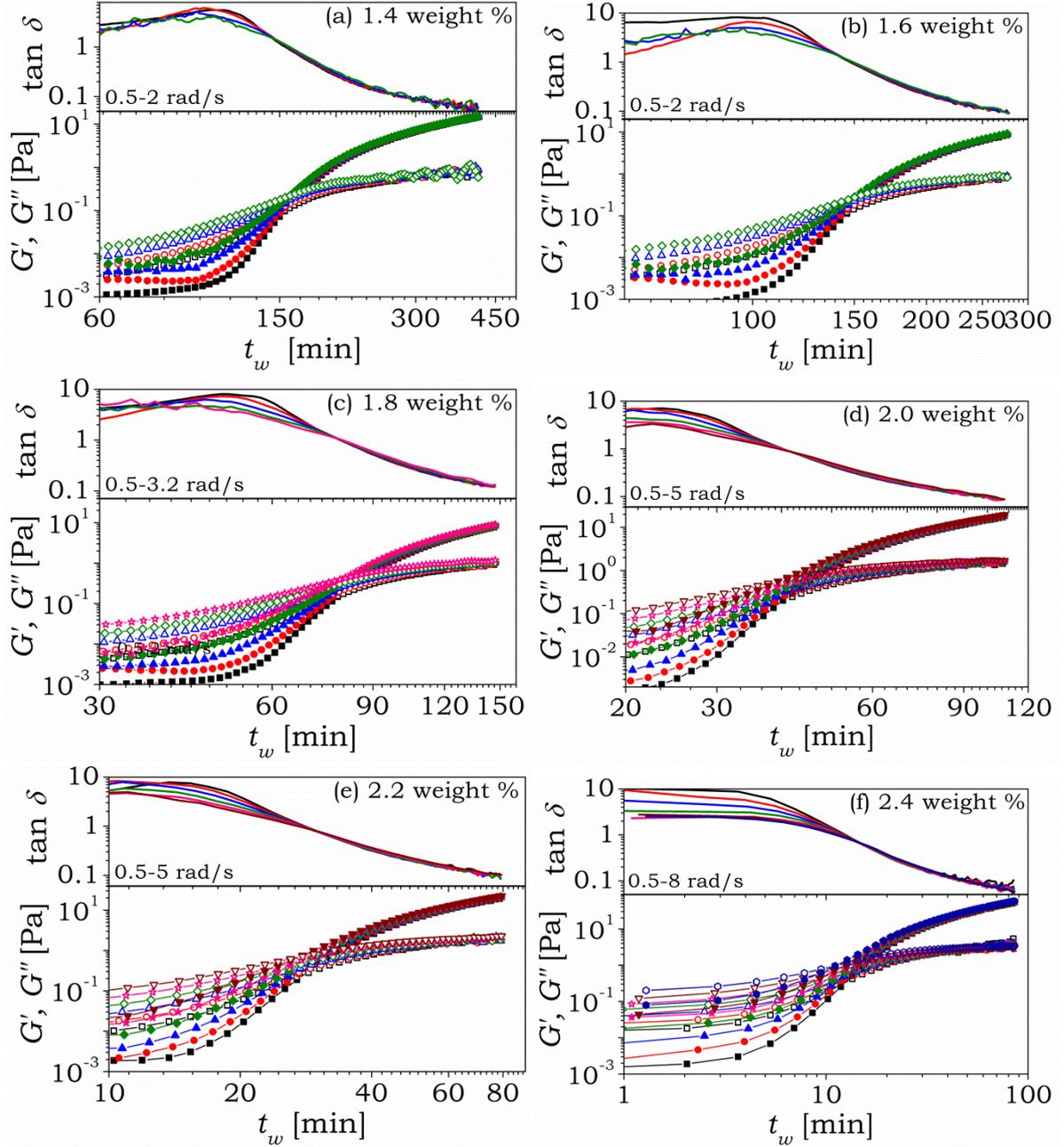

**Figure S8.** Viscoelastic properties $G'$ (filled symbols), $G''$ (open symbols) and $\tan\delta$ for suspensions of $C_s = 7$ mM and $C_L = 1.4$ (a), 1.6 (b), 1.8 (c), 2.0 (d), 2.2 (e), and 2.4 wt % (f) are plotted as a function of $t_w$ at different frequencies 0.5, 0.8, 1.25, 2, 3.2, 5, and 8 rad/s (for $G'$ and $G''$ from bottom to top and for $\tan\delta$ on the left hand side from top to bottom). Due to fluctuations, few frequencies data are not shown for better clarity.

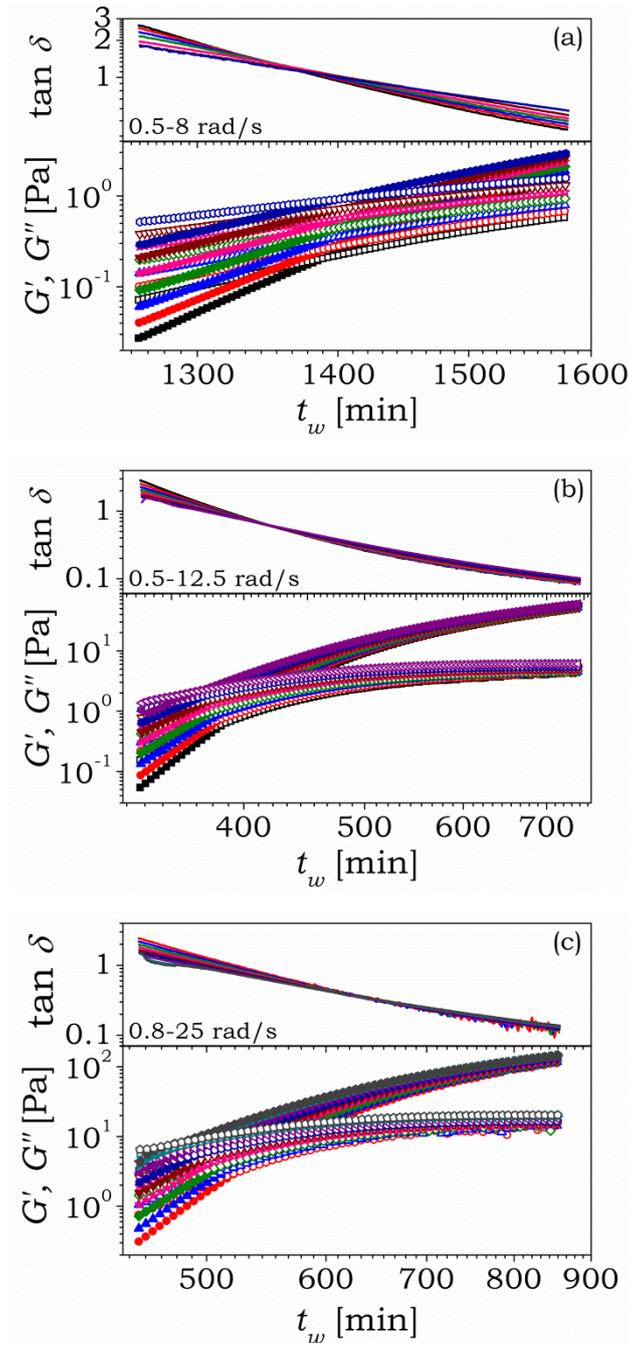

**Figure S9.** Viscoelastic properties $G'$ (filled symbols), $G''$ (open symbols) and $\tan\delta$ for filtered (0.45 $\mu$m filter) suspension samples of $C_L$=1.8 wt % with $C_s$=3 mM (a), $C_L$=2.4 wt % with $C_s$=3 mM (b), and $C_L$=3.4 wt % with $C_s$=0 mM (c) are plotted as a function of $t_w$ at different frequencies 0.5, 0.8, 1.25, 2, 3.2, 5, 8, 12.5, 20 and 25 rad/s (for $G'$ and $G''$ from bottom to top and for $\tan\delta$ on the left hand side from top to bottom). Due to fluctuations, few frequencies data are not shown for better clarity.

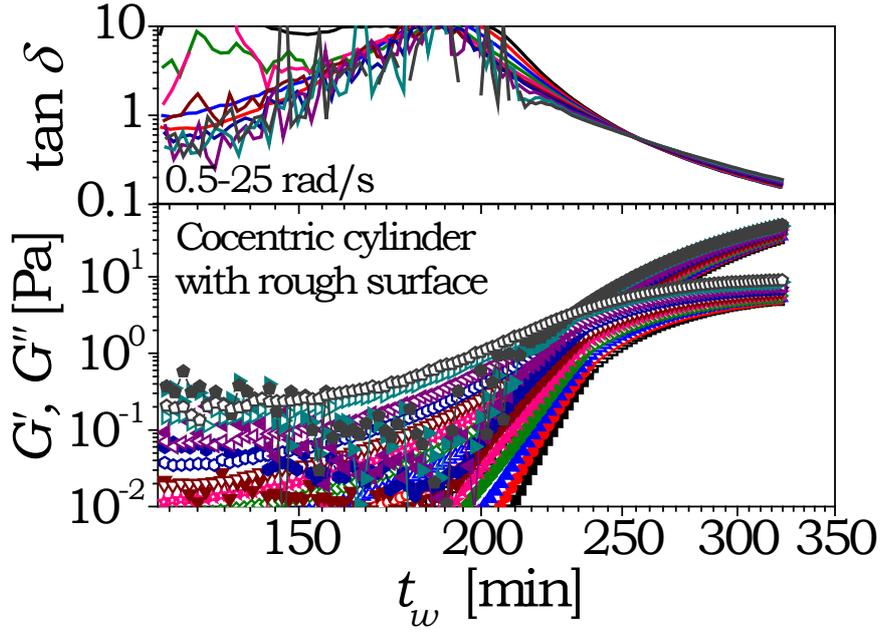

**Figure S10.** Viscoelastic properties $G'$ (filled symbols), $G''$ (open symbols) and $\tan\delta$ behavior in serrated couette geometry for $C_L$ =2.8 wt % with $C_S$ =3 mM suspension are plotted as a function of $t_w$ at different frequencies 0.5, 0.8, 1.25, 2, 3.2, 5, 8, 12.5, 20 and 25 rad/s (for $G'$ and $G''$ from bottom to top and for $\tan\delta$ on the left hand side from top to bottom).

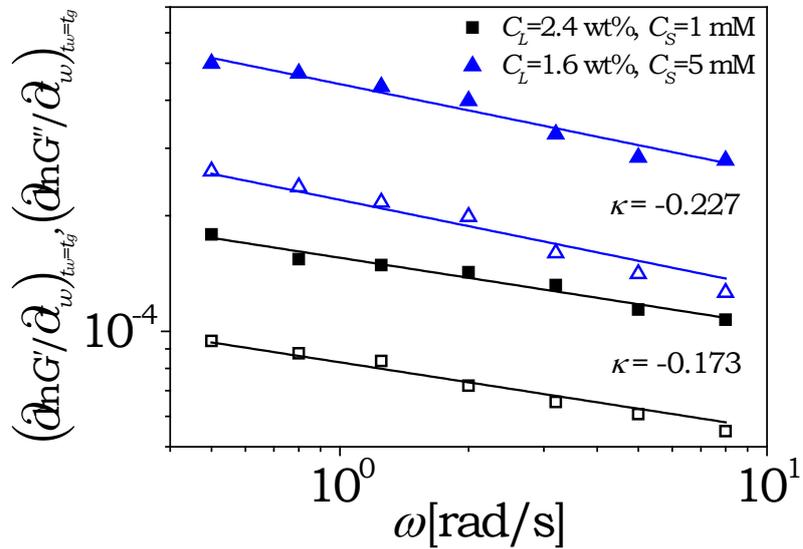

**Figure S11.** Growth rate of $G'$ (filled symbols) and $G''$ (open symbols) is plotted as a function of $\omega$ at the critical gel point for two Laponite suspension systems as mentioned in the label. The solid line represents a fit obtained of equation (6). The obtained values of dynamic critical exponent ($\kappa$) for both the suspensions are also mentioned.

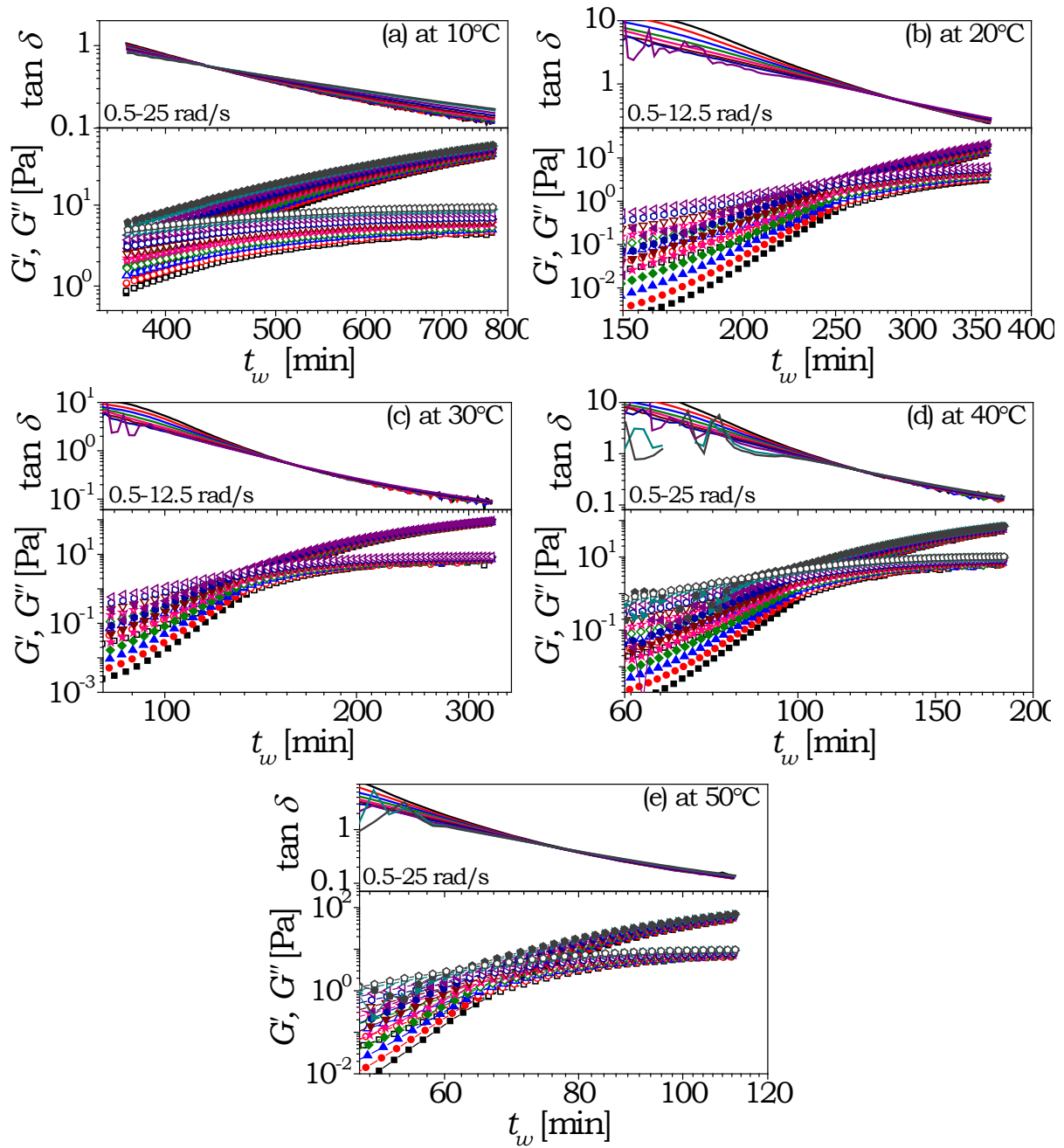

**Figure S12.** Viscoelastic properties $G'$ (filled symbols), $G''$ (open symbols) and $\tan \delta$ for suspensions of $C_L$ =2.8 wt % with $C_s$ =3 mM are plotted as a function of $t_w$ at temperature 10 (a), 20 (b), 30 (c), 40 (d) and 50 °C (e) at different frequencies 0.5, 0.8, 1.25, 2, 3.2, 5, 8, 12.5, 20 and 25 rad/s (for $G'$ and $G''$ from bottom to top and for $\tan \delta$ on the left hand side from top to bottom). Due to fluctuations, few frequencies data are not shown for better clarity.